\documentclass[prb,twocolumn,showpacs,aps,superscriptaddress,floatfix]{revtex4}
\usepackage{amsmath}
\usepackage{amssymb}
\usepackage{bm}
\usepackage{graphicx}
\usepackage{color}
\usepackage{hyperref}
\usepackage{verbatim}
\begin{document}
	\title{Thin film of a topological insulator as a spin Hall insulator}
	\author{R.S. Akzyanov}
		\affiliation{Dukhov Research Institute of Automatics, Moscow, 127055 Russia}
		\affiliation{Institute for Theoretical and Applied Electrodynamics, Russian
			Academy of Sciences, Moscow, 125412 Russia}
		\affiliation{P.N. Lebedev Physical Institute of the Russian
			Academy of Sciences, Moscow, 119991 Russia}

	\begin{abstract}
	We study the spin conductivity of the surface states in a thin film of a topological insulator within Kubo formalism. Hybridization between the different sides of the film opens a gap at the Dirac point. We found that in the gapped region spin conductivity remains finite. In the gapless region near the band gap spin conductivity is enhanced. These findings make a thin film of a topological insulator as a promising material for spintronic applications.
	\end{abstract}
	
	
	\maketitle
\section{Introduction} 
Spin Hall effect - generation of the transverse spin current by the applied voltage - have been predicted in the materials with spin-orbit scattering~\cite{Dyakonov1971} and with strong spin-orbit interaction~\cite{Murakami2003,Sinova2004}. However, it has been shown that spin current in latter materials is small due to vertex corrections caused by point non-magnetic disorder~\cite{Inoue2004,Raimondi2005}, that is consistent with the experimental data~\cite{Sinova2015}. 

While spin current is dissipationless itself~\cite{Murakami2003}, the accompanying charge current is dissipative. Ideal material for the spintronics should have high spin conductivity along with low charge conductivity. In fact, a finite spin current can be produced in the insulators, where the charge current is absent due to the band gap. Such effect is referred to quantum spin Hall effect (QSHE) and it is predicted in narrow gap semiconductors~\cite{Murakami2004}, graphene with enhanced spin-orbit interactions~\cite{Kane2005} and strained Rashba materials~\cite{Bernevig2006}. Also, QSHE is predicted in transition metal dichalcogenides, but for the experimentally relevant conditions spin Hall conductivity inside the gap does not occur~\cite{Li2013}. 

Topological insulators have robust surface states that form Dirac cone due to topologically nontrivial band structure in the bulk~\cite{Hasan2010}. Such materials have a potential in spintronics: record spin currents have been reported recently~\cite{Mellnik2014,Fan2014,Wang2015a,Fan2016,Khang2018}. In a thin film finite hybridization between surface states open a gap at the Dirac point~\cite{Lu2010,Shan2010}. We argue that finite spin conductivity exists in the gapped region of such film.


In the previous paper, we study the bulk and surface spin conductivity of the thick topological insulators~\cite{Akzyanov2019}. Also, spin conductivity in a thin film of a topological insulator is studied without taking into the account of the intralayer scattering and vertex corrections in Ref.~\cite{Peng2016}. The case of the gapped surface states due to intralayer hybridization has been missed. In this paper, we focus on the effects of the finite hybridization between the surface states in a thin film of a topological insulator. 

We calculate the spin conductivity of the surface states of a thin film of topological insulator using Kubo formulas taking into account vertex corrections. We found that spin conductivity remains finite in a gapped region. Spin conductivity is enhanced near the gap in a metallic region. These finding can open the road for the application of thin films of topological insulators in low dissipative spintronics.

\section{Model}
Low energy surface states in the thin film of topological insulator can be described by the Hamiltonian~\cite{PhysRevB.82.045122,Lu2010,Shan2010} ($\hbar=1$)
\begin{eqnarray}\label{H0}
\hat{H}\!=\!r(k_x^2+k_y^2)+\mu+v_{Fk}(k_x \sigma_y - k_y \sigma_x)\tau_z+\Delta \tau_x,\\
\nonumber
v_{Fk}=v_F[1+s(k_x^2+k_y^2)],
\end{eqnarray}
where $\mathbf{\sigma}=(\sigma_x,\sigma_y,\sigma_z)$ are the Pauli matrices acting in the spin space, $\mathbf{\tau}=(\tau_x,\tau_y,\tau_z)$ are the Pauli matrices acting in the layer space, $\mu$ is the chemical potential, $v_F$ is the Fermi velocity, $r=1/(2m)$ is the inverse mass term, $s$ characterizes the next order correction to the Fermi velocity, $k_x=k\cos \phi$ and $k_y=k\sin \phi$ are the in-plane momentum components, $\Delta$ is the value of the gap at the Dirac point due to hybridization of the surface states belong to different layers. The spectrum of the Hamiltonian~\eqref{H0} is doubly degenerate and is given by
\begin{eqnarray}\label{spectrum}
E_{\pm}=\mu +rk^2\pm
\sqrt{v_F^2k^2(1+sk^2)^2+\Delta^2}.
\end{eqnarray}
If we measure the energy in terms of $v_F^2/r$, then, the chemical potential, the next order correction to the Fermi velocity, and the gap are conveniently characterized by the dimensionless values $r\mu /v_F^2$, $sv_F^2/r^2$, and $r\Delta /v_F^2$, respectively.

The spectrum forms a Dirac cone when the inverse mass term is small $sv_F^2/r^2>1/3$. In the opposite case, an additional Fermi surface with opposite helicity emerges for some values of chemical potential~\cite{Akzyanov2019}. Note, that vanished inverse mass term $r=0$ leads to vanished spin conductivity. 

\begin{figure}[t!]
	\center
	\includegraphics [width=8.5cm, height=4.5cm]{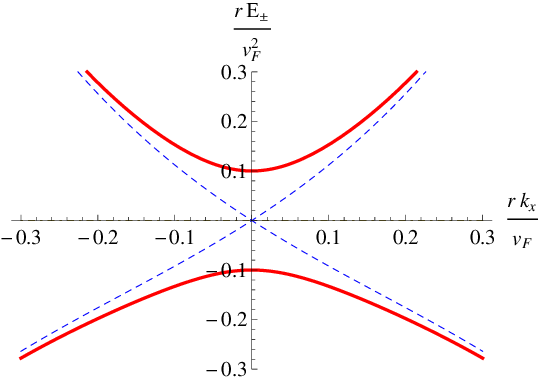}
	\caption{Energy spectrum~\eqref{spectrum} for different values of the gap $\Delta$ for $sv_F^2/r^2=2$. Blue dashed line is the spectrum for zero gap $\Delta=0$, red solid line is the spectrum for finite gap $r\Delta/v_F^2=0.5$.}
	\label{spectra}
\end{figure}

In general, the spin conductivity can be presented as a sum of three terms~\cite{Yang2006,Kodderitzsch2015}
\begin{eqnarray}
\sigma_{{\alpha}\beta}^{\gamma}=\sigma_{{\alpha}\beta}^{ I\gamma}+\sigma_{{\alpha}\beta}^{ II\gamma}+\sigma_{{\alpha}\beta}^{III\gamma },
\end{eqnarray}
where the first two items correspond to a contribution from the states at the Fermi level and the third one from the filled states. Here $\alpha$ and $\beta$ denote the in-plane coordinates $x$ and $y$ correspondingly, and $\gamma$ denotes the spin projection.

At zero temperature $\sigma_{{\alpha}\beta}^{I\gamma}$ and $\sigma_{{\alpha}\beta}^{II\gamma}$ contribution of the states at the Fermi energy can be written as~\cite{Inoue2004,Yang2006}
	\begin{eqnarray}\label{spin_surface}
\sigma_{{\alpha}\beta}^{I\gamma }=\frac {e}{4\pi}\int \frac{d^2k}{(2\pi)^2} \textrm{Tr} [j_{\alpha}^{\gamma}\, G^+
\,V_{\beta}  \, G^-] ,\\
\label{spin_surface1}
\sigma_{{\alpha}\beta}^{II\gamma }=-\frac {e}{8\pi}\int \frac{d^2k}{(2\pi)^2}\textrm{Tr} [j_{\alpha}^{\gamma}\, G^+ \,V_{\beta}  \, G^+ + j_{\alpha}^{\gamma}\, G^- \,V_{\beta}  \, G^- ].
\end{eqnarray}
Here $j_{\alpha}^{\gamma}= \{\sigma_{\gamma}, v_{\alpha}\}/4$,  $v_{\alpha}= \partial H /\partial k_{\alpha}$ is the velocity operator, $V_{\alpha}=v_{\alpha} + \delta v_{\alpha}$ is the velocity operator with vertex corrections $\delta v_{\alpha}$, $\{\,,\}$ means the anticommutator, and $G^{\pm}$ are the retarded and advanced disorder averaged Green functions, which will be specified later.

The contribution to the spin conductivity from the filled states 
\begin{eqnarray}\label{spin_filled}
\sigma_{{\alpha}\beta}^{III\gamma}=\frac {e}{8\pi}\int \frac{d^2k}{(2\pi)^2} \int\limits_{-\infty}^{\mu} f(E) dE \times\\ \nonumber \textrm{Tr} [j_{\alpha}^{\gamma}\, G^+\,v_{\beta}  \, \frac {d G^+}{dE}-j_{\alpha}^{\gamma}\, \frac {d G^+}{dE}\,v_{\beta}  \,G^+ + c.c.] ,
\end{eqnarray}
where $f(E)$ is the Fermi distrubution function (that is Heaviside step for zero temperature), $c.c$ means complex conjugate.

\section{Disorder and vertex corrections}
We will describe disorder by a potential $V_{\textrm{imp}}=u_0 \sum\limits_i \delta(\mathbf{r}-\mathbf{R}_j)$, where $\delta(\mathbf{r})$ is the Dirac delta function, $\mathbf{R}_j$ are the positions of the randomly distributed point-like impurities with the local potential $u_0$ and concentration $n_i$. We assume that the disorder is Gaussian, that is, $\langle V_{\textrm{imp}} \rangle=0$ and $\langle V_{\textrm{imp}}(\mathbf{r}_1) V_{imp}(\mathbf{r}_2) \rangle=n_i u_0^2 \delta (\mathbf{r}_1-\mathbf{r}_2)$. We introduce disorder parameter as $\gamma_b=n_iu_0^2/(4v_F^2)$.

In the self-consistent Born approximation (SCBA), the impurity-averaged Green's functions can be calculated as
$
G^{\pm}=G_0^{\pm}+G_0^{\pm}\Sigma^{\pm} G^{\pm},
$
where $G_0^{\pm}$ are bare retarded/advanced Green's functions of the Hamiltonian~\eqref{H0}
and $\Sigma^{\pm}$ is the self-energy. Self-energy is defined as
$
\Sigma^{\pm}=  \langle V_{\textrm{imp}} G^{\pm} V_{\textrm{imp}}\rangle.
$
In the case under consideration, we can calculate the self-energy $\Sigma^{\pm}=\Sigma'\mp i\Gamma$ using an Dyson equation $\Sigma^{\pm}=n_iu_0^2\sum_k G^{\pm} $.	
The self-energy has nontrivial structure in the side space $\tau$. Along with diagonal element $\Sigma_0 \tau_0$ it has non-diagonal one $\Sigma_x \tau_x$. Therefore, the expression for $G^\pm$ is similar to $G^\pm_0$, in which $\pm i0$ is replaced by $\pm i\Gamma_0$, $\mu$ by $\mu-\Sigma'_0$ and $\Delta$ by $\Delta-\Sigma'_x+i\Gamma_x$ . The value $\Gamma_0$ describes diagonal scattering rate while $\Gamma_x$ describes non-diagonal scattering rate.

We start with the case of large chemical potential $|\mu| \gg \Delta$. In this case we can neglect a small correction to the value of $\mu$ due to real part of the self-energy and put $\Sigma' =0$. In this limit we suppose that scattering rates $\Gamma_0, \Gamma_x \rightarrow 0$ are small and we obtain that
$
\Gamma=\textrm{Im}\Sigma^{+} =n_i u_0^2\sum_k\,\textrm{Im} G_0^+.
$

Now we consider $r=s=0$.  In the metallic region $|\mu|>\Delta$ we get that diagonal scattering rate is independent of chemical potential~\cite{Kato2007}
\begin{eqnarray}\label{scatter}
\Gamma_0=-\gamma_b |\mu| , \quad
\Gamma_x=\gamma_b\Delta \mu/|\mu|
\end{eqnarray}
Condition $\Gamma_0 \Delta = -\Gamma_x \mu$ stands even for $|\mu| \succeq \Delta$.
Near the band gap $|\mu| \simeq \Delta$ we calculate scattering rates self-consistenly and found that scattering rates are exponentially suppressed for a weak disorder $\Gamma_{0(x)} \propto e^{-2/(\pi \gamma_b)}$ that is expected for the Dirac system~\cite{Ostrovsky2006}. In the insulating region $|\mu|<\Delta$ condition Eq.~\ref{scatter} fails and non-diagonal scattering is always smaller than diagonal one $|\Gamma_0|>|\Gamma_x|$.

The impurity averaged Green function can be calculated as $G^{\pm}=(1+\Sigma G_0^{\pm})^{-1} G_0^{\pm}$ or in the explicit form
	\begin{eqnarray}\label{greenF0}
	G^{\pm}\!=\!\frac{\mu\!+\!rk^2\!\pm\! i\Gamma_0\!-\!v_{Fk}(k_x \sigma_y\! -\! k_y \sigma_x)\tau_z \!-\!  (\Delta\!\pm\!i\Gamma_x)\tau_x\!}{(\mu+rk^2\pm i\Gamma_0)^2-\left[v_{Fk}^2k^2+(\Delta\pm i\Gamma_x)^2\right]}\,.
	\end{eqnarray}

In the SCBA, following the approach described in Ref.~\cite{Shon1998}, we can derive an equation for the vertex corrected velocity operator~\cite{Chiba2017}
\begin{eqnarray}\label{vcor}
V_{\alpha}(\mathbf{k})=v_{\alpha}(\mathbf{k}) + \frac{n_i u_0^2}{(2\pi)^2} \int G^+(\mathbf{k}) V_{\alpha}(\mathbf{k}) G^-(\mathbf{k})d^2\mathbf{k}.
\end{eqnarray}

We found that hybridization $\Delta$ has a little influence on the vertex corrections. For $r=s=0$ we get standart expression for the vertex corrected velocity operator for the Dirac spectrum $V_{\alpha}=2v_{\alpha}$~\cite{Ostrovsky2006,Chiba2017}. 

Point-like disorder renormalize $k$-independent part of the velocity operator. Thus, we write down
\begin{eqnarray}\label{hrz}  
V_x=v_x + \delta v\sigma_y\tau_z, \\ \nonumber
V_y=v_y- \delta v\sigma_x\tau_z,
\end{eqnarray}
where vertex corrections $\delta v$ are calculated by the substitution of Eq.~\eqref{hrz} into Eq.~\eqref{vcor}.  

\section{Spin conductivity from the states at the Fermi surface}
Now we use the obtained results and Eqs.~\eqref{spin_surface} to calculate the contribution to the spin conductivity due to the states at the Fermi surface. On this way, we obtained, first, that in the considered approach the term $\sigma_{{\alpha}\beta}^{II\gamma}$ vanishes exactly and we should compute the term $\sigma_{{\alpha}\beta}^{I\gamma}$ only.

Isotropic tensor component $\sigma_{xy}^{Iz}=-\sigma_{yx}^{Iz}$ is the only term that exists in the system.	
From Eq.~\eqref{spin_surface} using condition $\Gamma_0 \Delta = \Gamma_x \mu$ we derive that 
	\begin{eqnarray}\label{spin1}
	\sigma_{xy}^{Iz}=\sigma_{xy}^{Izbv} + \delta \sigma_{xy}^{Iz}, \\ \nonumber
	\sigma_{xy}^{Izbv}=-\sigma_0^z \int k\,dk\, \frac {8r\Gamma_0 v_{Fk}^2 k^2 }{\pi E_{g+}E_{g-}},  \\ \nonumber
\delta	\sigma_{xy}^{Iz}=-\sigma_0^z \int k\,dk\, \frac {8r\Gamma_0 v_{Fk}\delta v k^2 }{\pi E_{g+}E_{g-}},  \\ \nonumber
	E_{g\pm}\!=\!(\mu+rk^2\pm i\Gamma_0)^2\!-\!v_{Fk}^2k^2-\left(\Delta\pm i\Gamma_x\right)^2.
	\end{eqnarray}
Here $\sigma_{xybv}^{Iz}$ is the spin conductivity without vertex corrections, $\delta \sigma_{xy}^{Iz}$ is the spin conductivity cotribution from the vertex corrections. 
We plot spin conductivity contributions as a function of the chemical potential for different values of the gap $\Delta$ and correction to the Fermi velocity $s$. We can see that spin conductivity contributions take maximal values near the gap $\mu \sim \Delta$. 
\begin{figure}[t!]\label{surface1}
	\center
	\includegraphics[width=8.5cm, height=4.5cm]{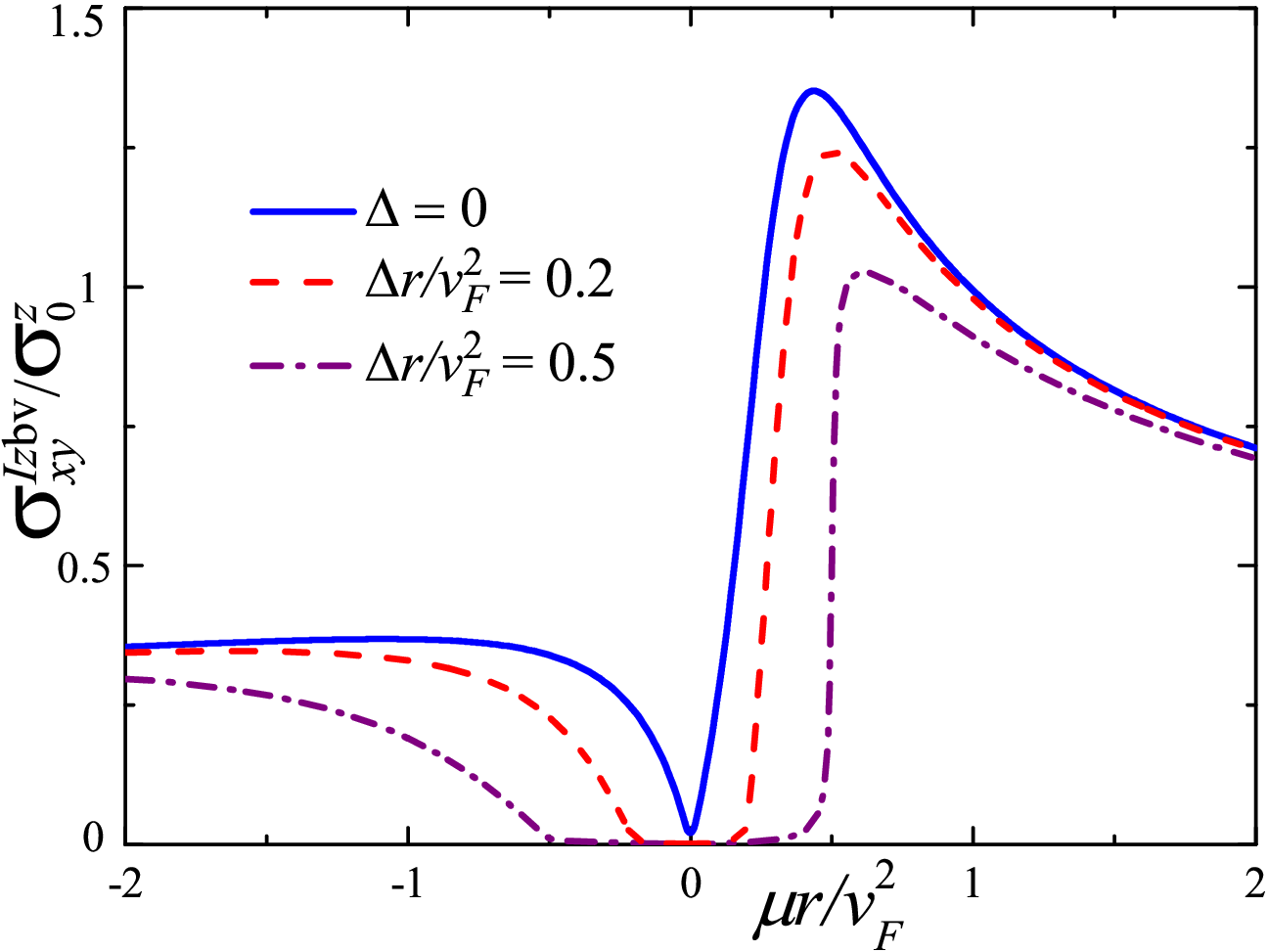}
	\includegraphics [width=8.5cm, height=4.5cm]{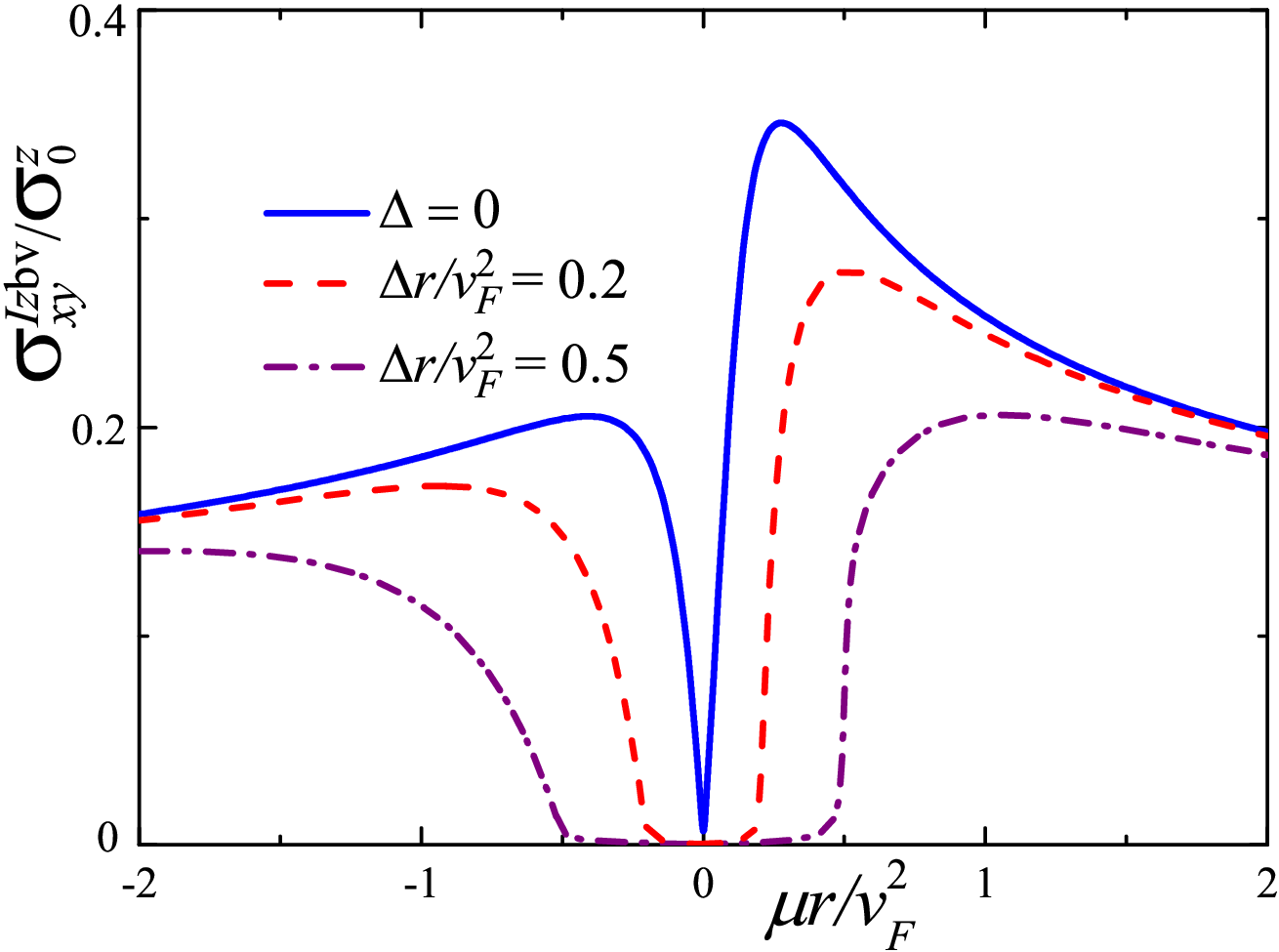}
	\caption{Bare spin conductivity $\sigma_{xybv}^{Iz}$ for $\gamma_b=0.001$, $sv_F^2/r^2=1$ (upper panel) and $sv_F^2/r^2=5$ (lower panel). Blue line corresponds to $\Delta=0$, red dashed line to $r\Delta/v_F^2=0.2$, purple dot dashed line to $r\Delta/v_F^2=0.5$.}
\end{figure}
\begin{figure}[t!]\label{surfacevc}
	\center
	\includegraphics [width=8.5cm, height=4.5cm]{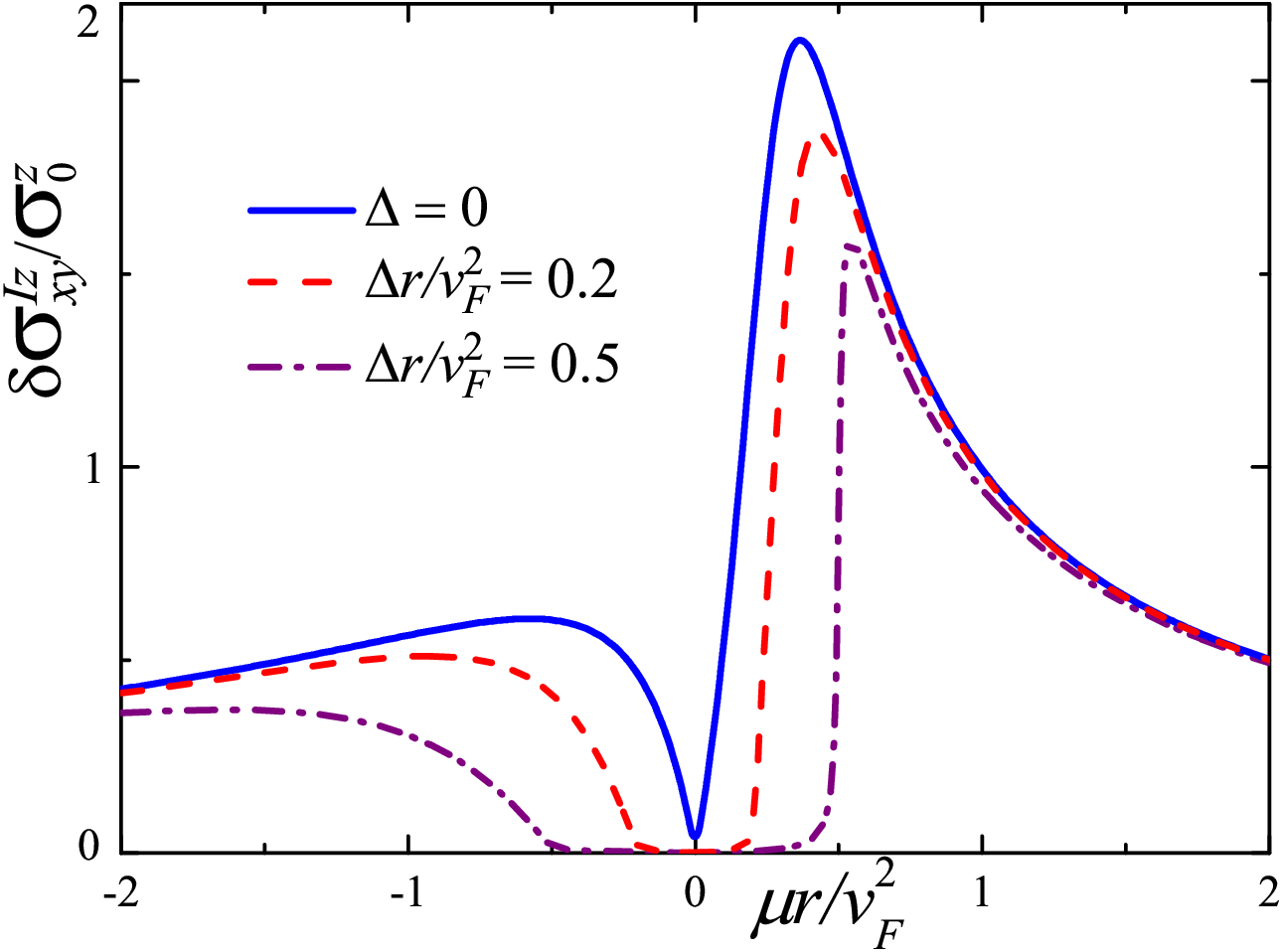}
	\includegraphics [width=8.5cm, height=4.5cm]{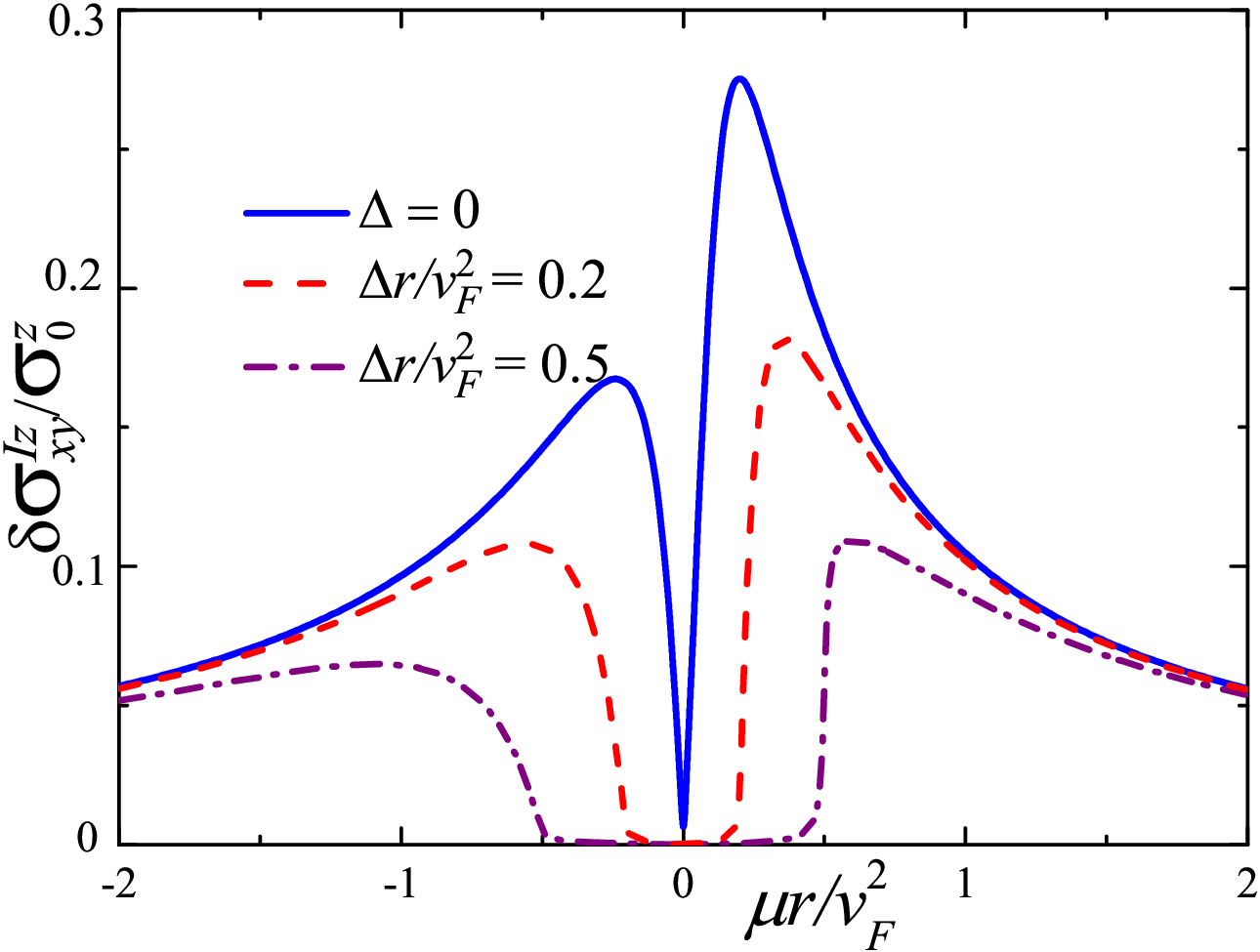}
	\caption{Spin conductivity contribution from vertex corrections $\delta \sigma_{xy}^{Iz}$ for $\gamma_b=0.001$, $sv_F^2/r^2=1$ (upper panel) and $sv_F^2/r^2=5$ (lower panel). Blue line corresponds to $\Delta=0$, red dashed line to $r\Delta/v_F^2=0.2$, purple dot dashed line to $r\Delta/v_F^2=0.5$.}
\end{figure}
\section{Spin conductivity from the filled states}
\begin{figure}[t!]
	\center
	\includegraphics [width=8.5cm, height=4.5cm]{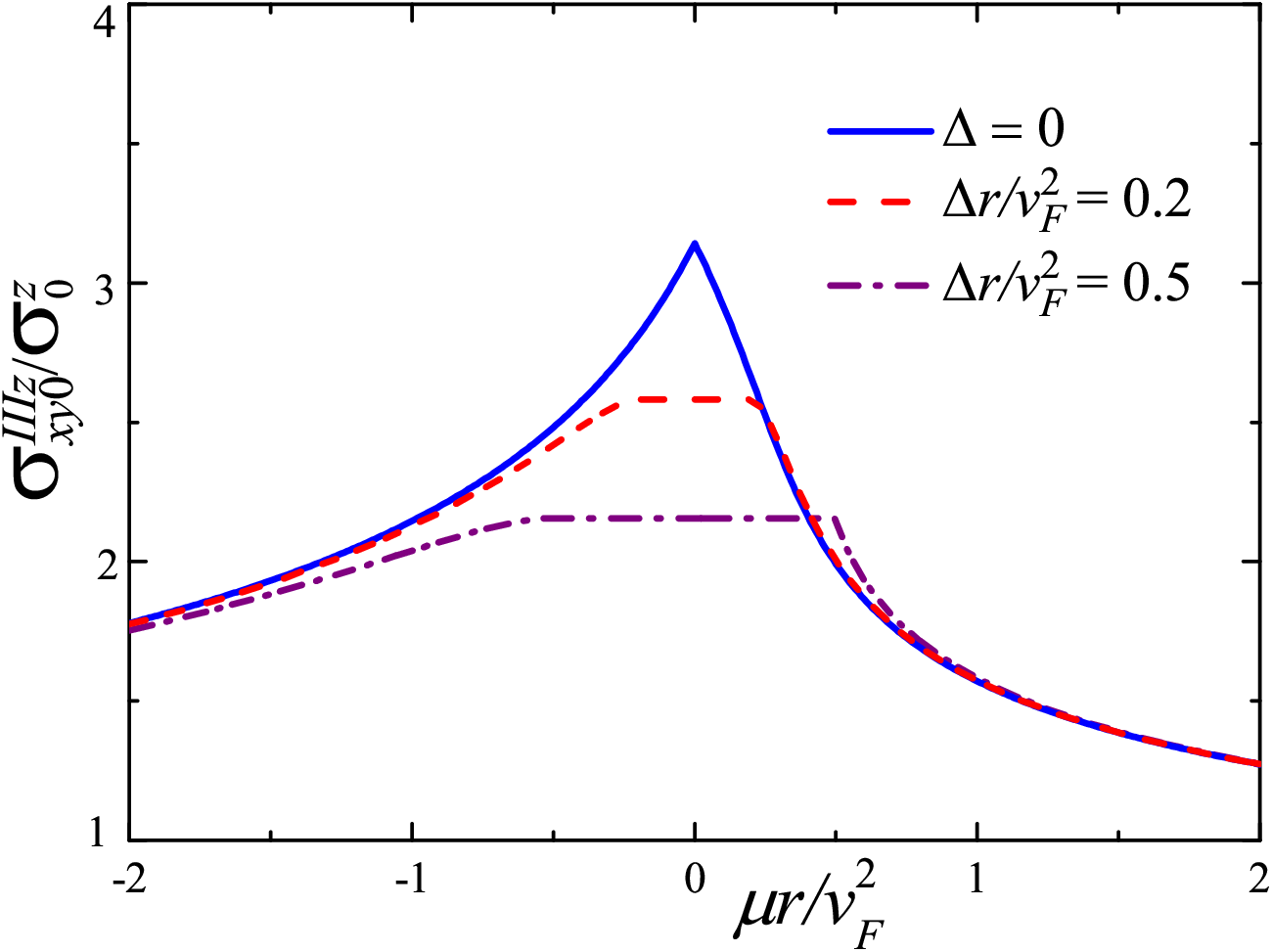}
	\includegraphics [width=8.4cm, height=4.5cm]{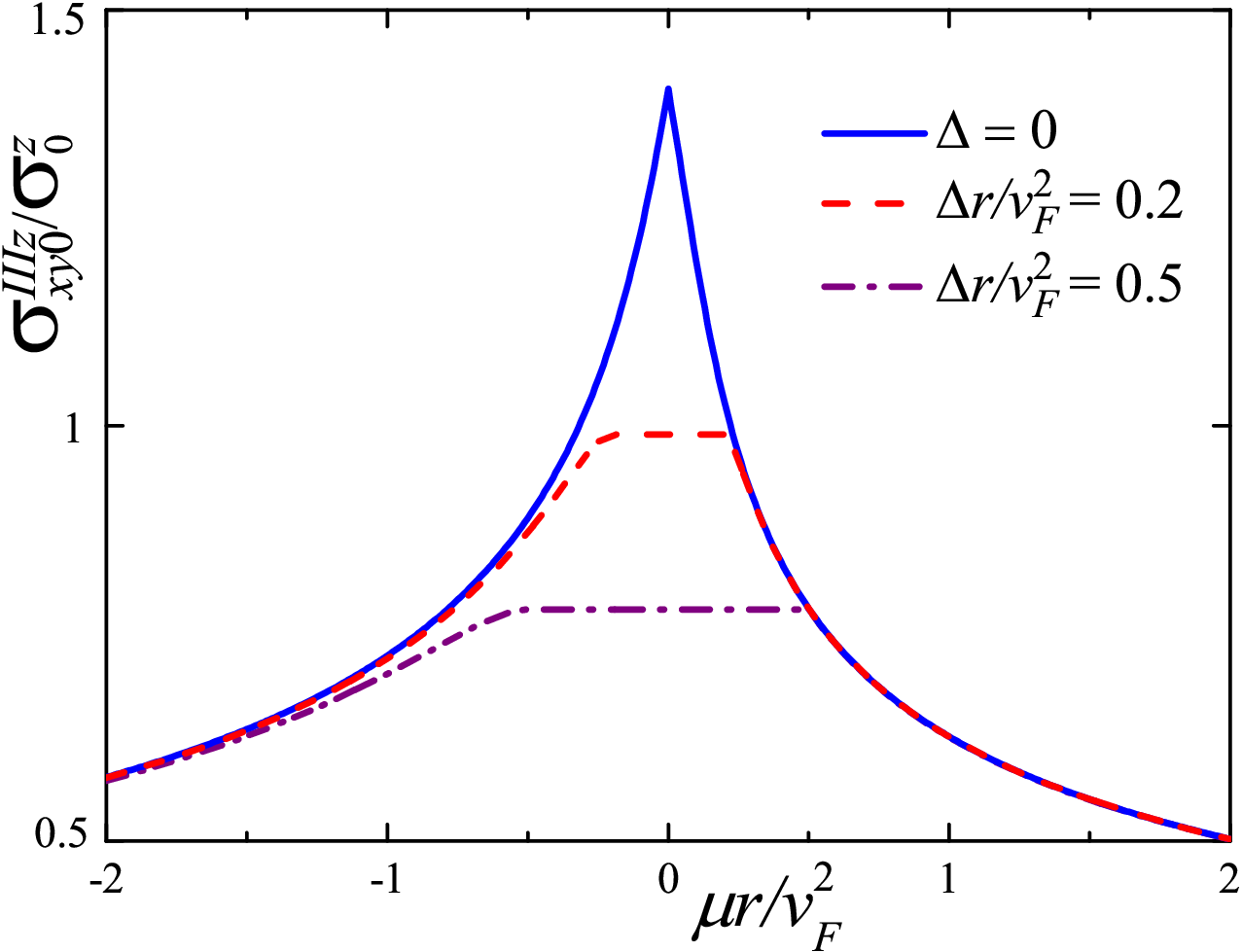}
	\caption{Spin conductivity in a clean limit $\sigma_{xy0}^{IIIz}$ as a function of chemical potential for $sv_F^2/r^2=1$ (upper panel) and for $sv_F^2/r^2=5$ (lower panel). Blue line corresponds to $\Delta=0$, red dashed line to $r\Delta/v_F^2=0.1$, purple dotdashed line to $r\Delta/v_F^2=0.5$, green line to $r\Delta/v_F^2=1$.}\label{top1}
\end{figure}

We study Eq.~\ref{spin_filled} in a weak disorder limit. Isotropic component $\sigma_{xy}^{IIIz}=-\sigma_{yx}^{IIIz}$ is the only term that exists in the system. After energy integration using constant damp apprxoimation for the $\Gamma_0$ we get that contribution to the spin conductivity from the filled states consists of two parts
\begin{equation} \label{spin2}
\sigma_{xy}^{IIIz}=\sigma_{xy0}^{IIIz}-\sigma_{xy}^{Izbv},
\end{equation}
where $\sigma_{xy0}^{IIIz}$ is the spin conductivity in a clean limit, $\sigma_{xy}^{Izbv}$ is given by Eq.~\ref{spin1}. Similar expression have been obtain for the anomalous charge Hall conductivity for the Dirac Hamiltonian~\cite{Sinitsyn2007}.
Spin conductivity from the filled states in a clean limit can be expressed as~\cite{Sinova2004,Sinitsyn2004}
\begin{eqnarray}\label{sigmaIII}
\sigma_{{\alpha}\beta 0}^{III \gamma}\!\!=\!\!e\!\!\!\! \sum\limits_{\mathbf{k},n\neq n'}\!\!\!(f_{n\mathbf{k}}\!-\!f_{n'\mathbf{k}})\!
\frac{ \textrm{Im}\langle u_{n'\mathbf{k}}|j_{\alpha}^{\gamma}|u_{n\mathbf{k}}\rangle\!\langle u_{n\mathbf{k}}|v_{\beta}|u_{n'\mathbf{k}}\rangle}{(E_{n\mathbf{k}}-E_{n'\mathbf{k}})^2}
\end{eqnarray}
Here $E_{n\mathbf{k}}$ is the energy of an electron in the $n$-th band with the momentum $\mathbf{k}$, $u_{n\mathbf{k}}$ is the corresponding Bloch vector, $\hat{H} u_{n\mathbf{k}} = E_{n\mathbf{k}} u_{n\mathbf{k}}$, $f_{n\mathbf{k}}$ is the Fermi distribution function corresponding to $E_{n\mathbf{k}}$ (which is the Heaviside step-function in the considered case of zero temperature), Im is for the imaginary part, $\langle ... \rangle$ is the scalar product here.
Using Eq.~\eqref{sigmaIII} we obtain 
\begin{eqnarray}
\sigma_{xy0}^{IIIz}=\sigma_0^z \int\limits_0^{\infty} (\theta(E_+)-\theta(E_-))k\,dk \frac {2rk^2 v_{Fk}^2}{(v_{Fk}^2+\Delta^2)^\frac{3}{2}}, 
\end{eqnarray}
where $\theta(x)$ is Heaviside step function and $\sigma_0^z=e/(8\pi)$ is the spin conductivity quanta. Finite disorder has a little impact on this term in the spin conductivity: Heaviside step function $\theta(E_{\pm})$ is replaced by the normalized arctangent function $2/\pi\arctan E_{\pm}/\Gamma_0$. This substitution leads to the insignificant blurring of the spin conductivity near the gap $\mu \simeq \pm \Delta$ for a small disorder.
Spin conductivity in a clean limit $\sigma_{xy0}^{IIIz}$ is shown on the Fig.~\ref{top1} for different values the the gap $\Delta$ and Fermi velocity correction $s$. Spin conductivity is a constant in the gapped region and decreases in the gapless region. Also, its particle-hole asymmetry is controlled by the parameter $sv_F^2/r^2$: asymmetry is smaller for larger values of the parameter. We can see that spin conductivity in a gapped region is comparable to the spin conductivity in a metallic region near the gap $\sigma_{xy0}^{IIIz}(\mu=0,\Delta) \simeq \sigma_{xy0}^{IIIz}(\mu=\Delta,\Delta=0)$, so at the Dirac point spin conductivity decreases with the increase of the gap $\Delta$.

\section{Estimates for the experiment}
We can extract information of the disorder from half-width of the quasiparticle peak in ARPES from Ref.\onlinecite{Chen2013} and get $\gamma_b \sim 10^{-2}$. STM study shows that  there is approximately 1 defect per $\AA^2$ for a clean surface~\cite{Cheng2010}. If we suppose that the typical impurity potential is is order of chemical potential $\mu \sim 200$cmeV (which is true for the vacation defects) and Fermi velocity for the surface states~\cite{Zhang2009} $v_F \sim 3$ eV$\cdot\AA^{-1}$, then we get estimate for $\gamma_b \sim 10^{-3}$.
Value of correction to the Fermi velocity $s$ for the surface states is extracted from the Ref.~\onlinecite{Rakyta2012}, and we get $sv_F^2/r^2=0.7$ where~\cite{Nomura2014} $v_F^2/r=1$eV. Hybridization between the layers $\Delta$ depends strongly on the number of layers and reaches values of $\Delta=0.2$eV for two layers of Bi$_2$Te$_3$~\cite{Forster2016}. 

We can see from Eqs.~\ref{spin1} and~\ref{spin2} that the term $\sigma_{xy0}^{Iz}$ cancels out from $\sigma_{xy}^{Iz}$ and $\sigma_{xy}^{IIIz}$. So, only contributions to the spin conductivity from the vertex corrections and spin conductivity in a clean limit remain. 
We plot total spin conductivity $\sigma_{xy}^{z}=\delta \sigma_{xy}^{Iz}+\sigma_{xy0}^{IIIz}$  as a function of chemical potential for the experimentally relevant parameters. Spin conductivity remains finite inside the gap and is enhanced just outside of the gap. Also, it posses considerable particle-hole asymmetry. Charge conductivity is calculated by~\cite{Proskurin2015} $\sigma_{xx} = \sigma_0/2 \sum_{\bf k} V_x (G^+-G^-)v_x(G^+-G^-)$, where $\sigma_0=e^2/{\hbar}$. Charge conductivity is suppressed inside the gap and it has a weak dependence of the value of chemical potential away from the gap as it is expected for the  Dirac system~\cite{Ostrovsky2006} 	
\begin{figure}[t!]\label{exp}
	\center
	\includegraphics [width=8.5cm, height=4.5cm]{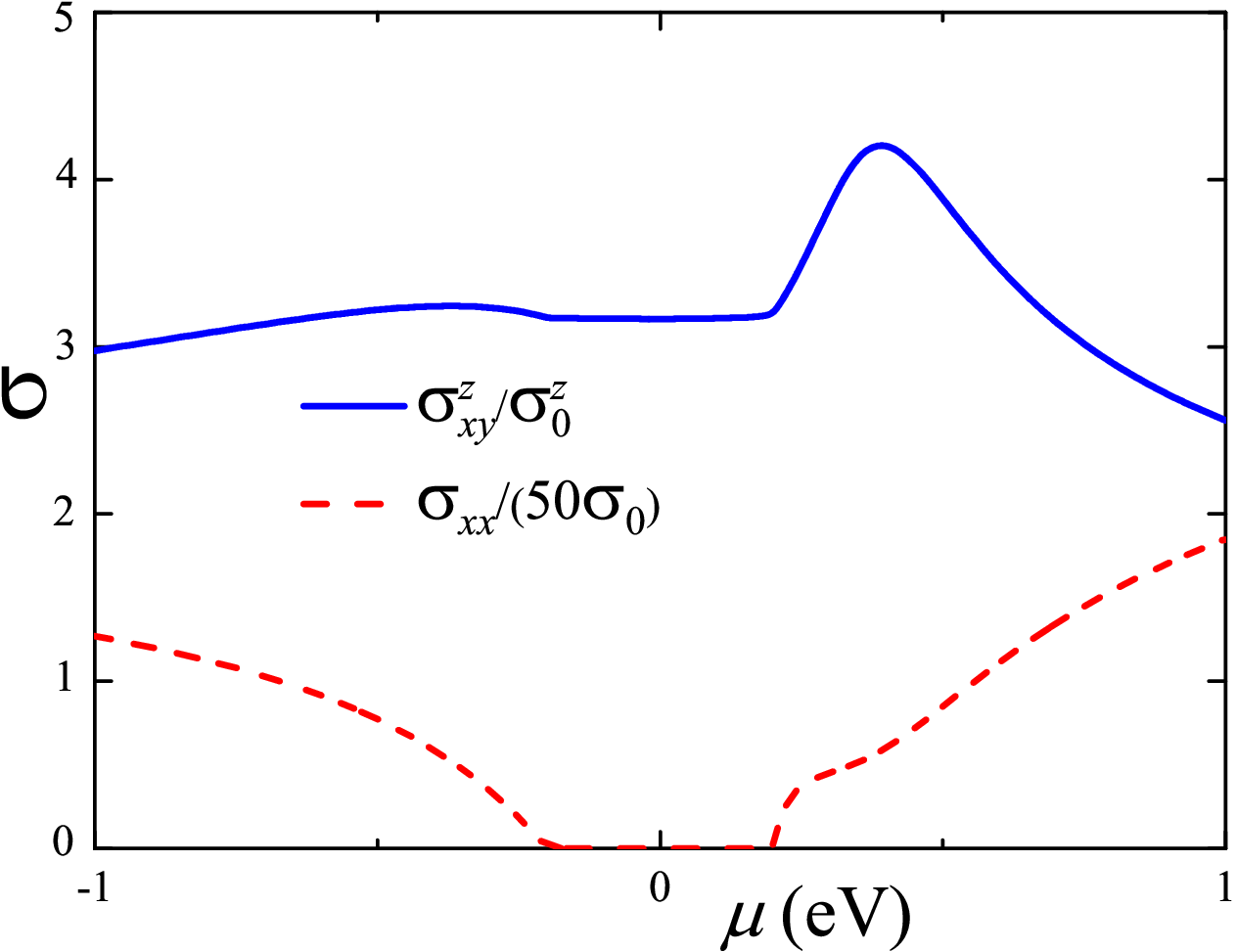}
	\caption{Total spin conductivity $\sigma_{xy}^{z}$ (blue line) and charge conductivity $\sigma_{xx}$ (dashed red line) for the experimentally relevant data $\gamma_b=0.01$, $sv_F^2/r^2=0.7$.}
\end{figure}

\section{Discussion}
Spin conductivity exists in a gapped region of a topological insulator thin film so this system is a spin Hall insulator. Spin conductivity is not changing inside the gap if we vary chemical potential. However, its value is not equal to the spin conductivity quanta because Hamiltonian does not commute with the z-component of the spin $[H,\sigma_z]\neq0$\cite{Dayi2015}, so spin is not conserved. It differs our system from the spin Hall insulators that have been considered before~\cite{Murakami2004,Kane2005,Bernevig2006,Li2013}. In a similar way charge quantum Hall conductivity is not quantized in superconductors since charge is not conserved in such materials~\cite{Bednik2016}.

In the Kane-Mele like models, where spin is conserved $[H,\sigma_z]=0$, spin conductivity is generated by the spinful interaction between the Dirac cones $\sim\sigma_z$~\cite{Kane2005,Li2013}. In our model interaction between the Dirac cones is independent of the spin and spin conductivity is generated by the intrinsic spin-orbit coupling as it is expected for the Rashba systems~\cite{Sinova2004}.

We suppose that both sides of the film are identical. In real samples asymmetry between different sides of the film can arise due to different doping and concentration of the defects. In a weak scattering limit, different scattering on the different sides of the film has almost no influence on the total spin conductivity if vertex corrections are neglected. However, significant asymmetry between the layers can have an effect on the vertex corrections. 

 Hybridization between layers, given by $\Delta \tau_x$, couples electron (hole) from the Dirac cone of one layer to the hole (electron) of Dirac cone of another layer with the same spin. Such a bound state does not carry a charge but carries double spin and has spin-momentum locking of the parent electron. If we apply the electric field, then the flow of such bound states with zero charges will produce spin current without the generation of the electric current. 

The region inside the gap is of special interest. Scattering and dissipation are strongly suppressed due to an absence of the states at the Fermi energy. However, finite spin conductivity is present. Such a phase can be promising for low dissipation spintronics applications.

Outside the gap spin conductivity is enhanced due to presence of the contribution of the states at the Fermi energy. This enhancement is significant only for a thin films $\Delta \sim v_F^2/r$. If there are several layers then the gap is significantly suppressed in comparison with the mass parameter $\Delta \ll v_F^2/r \sim 1$eV. For the 4 layer film, bandgap reaches a value of $\Delta\sim180$meV and is almost vanished for thicker samples~\cite{Forster2016}. Thus, the effect of the enhancement of spin conductivity is significant only in a few layers samples. 

Experimentally, the dissipationless spin current can be measured by the spin-transfer torque effect~\cite{Ralph2008,Sinova2015}. If we apply voltage, then magnetization of the magnetic layer on the top of the topological insulator can be changed by the spin current without the generation of the electrical current. Thus, the ratio of spin current to charge current - spin Hall angle - would be extremely large for such a system. However, the magnetic layer on the top should be an insulator to prevent shunting.

We do not consider the influence of the magnetic field on the spin conductivity. In the real experiments on spin-transfer torque magnetic layers are used. Thus, the magnetization of the magnetic layer affects the value of the spin conductivity of the topological insulator. However, if magnetization is small (value of the Zeeman splitting is much smaller then chemical potential $|B|\ll |\mu|$) then its influence on the spin conductivity can be neglected.  

Experiments on spin-transfer torque in a film of topological insulators with eight layers have been performed~\cite{Mellnik2014}. Good quality ultrathin films of topological insulators with a few layers are available\cite{Zhang2009a,Kim2013}. The gap for the surface states is about 180 meV for the four-layer samples~\cite{Lang2013}. So, the measure of spin current in a thin layer topological insulator inside and near the gap is as an experimentally achievable task. 

To sum up,
hybridization between the surface states in different layers of a thin film of a topological insulator opens a gap near the Dirac point. We found that finite spin conductivity exists in the gapped region. In a metallic region near the gap, spin conductivity is enhanced. These findings can be crucial towards the implementation of thin films of topological insulators in low-dissipation spintronics.	

\section*{Acknowledgements}

We acknowledge support from the Russian Science Foundation, Grant No 17-12-01544, RFBR according to the research project № 19-32-80014, Foundation for the Advancement of Theoretical Physics and Mathematics “BASIS”.

	\bibliographystyle{apsrevlong_no_issn_url}
	\bibliography{bib_hw}

\end{document}